\documentclass{article}

\usepackage{graphicx}
\usepackage{arxiv}
\usepackage[utf8]{inputenc}
\usepackage[T1]{fontenc}
\usepackage{amsmath,amssymb}
\usepackage{amsthm}
\usepackage{booktabs}
\usepackage{tabularx}
\usepackage{array}
\usepackage{microtype}
\usepackage{url}
\usepackage{listings}
\usepackage{xcolor}
\usepackage{multirow}
\usepackage{pifont}
\usepackage[colorlinks=true,urlcolor=blue,citecolor=black,
            anchorcolor=black,linkcolor=black]{hyperref}

\newtheorem{definition}{Definition}

\newcolumntype{L}{>{\raggedright\arraybackslash}X}
\newcolumntype{C}{>{\centering\arraybackslash}X}
\newcolumntype{R}{>{\raggedleft\arraybackslash}X}

\newcommand{\mad}{\textsc{mad}}   % metric aggregation divergence shorthand
\newcommand{\woc}{WoC}            % Without Contract
\newcommand{\wc}{WC}              % With Contract

\title{Metric Aggregation Divergence: A Hidden Validity Threat in\\
       Agent-Based Policy Optimization and a Contractual Remedy}

\author{
\href{https://orcid.org/0000-0002-0883-4574}{\includegraphics[scale=0.06]{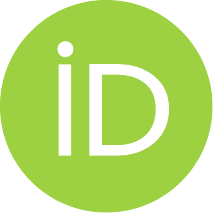}\hspace{1mm}\textcolor{black}{Ruiyu Zhang}}\\
Department of Politics and Public Administration\\
The University of Hong Kong\\
\texttt{ruiyuzh@connect.hku.hk}\\
\And
\href{https://orcid.org/0000-0002-0275-117X}{\includegraphics[scale=0.06]{orcid.pdf}\hspace{1mm}\textcolor{black}{Lin Nie}}\\
Department of Applied Social Sciences\\
The Hong Kong Polytechnic University\\
\texttt{lin-apss.nie@polyu.edu.hk}\\
\And
\href{https://orcid.org/0009-0005-7399-109X}{\includegraphics[scale=0.06]{orcid.pdf}\hspace{1mm}\textcolor{black}{Xin Zhao}}\\
Department of Applied Social Sciences\\
The Hong Kong Polytechnic University\\
\texttt{xinnn.zhao@connect.polyu.hk}
}

\renewcommand{\headeright}{A Preprint}
\renewcommand{\undertitle}{A Preprint}
\renewcommand{\shorttitle}{Metric Aggregation Divergence}

\hypersetup{
    pdftitle={Metric Aggregation Divergence: A Hidden Validity Threat in Agent-Based Policy Optimization and a Contractual Remedy},
    pdfauthor={Ruiyu Zhang, Lin Nie, Xin Zhao},
    pdfkeywords={agent-based modeling, policy optimization, multi-objective evolutionary algorithms, validity threats, metric aggregation, reproducibility, NSGA-II},
}

\begin{document}

\maketitle

% ============================================================
% ABSTRACT
% ============================================================
\begin{abstract}
Metric aggregation divergence (MAD) is the silent inconsistency that arises when distinct pipeline stages in an agent-based model coupled with a multi-objective evolutionary algorithm (ABM+MOEA) independently re-implement how an outcome metric is extracted from simulation trajectories. Unlike deliberate analysis choices, MAD operates at the level of pipeline architecture: each stage is internally coherent, and the inconsistency becomes visible only when cross-stage outputs are compared. Code inspection of \emph{EpidemiOptim}, a JAIR-published epidemic policy toolbox, reveals three structurally independent aggregation paths in peer-reviewed code. A faithful replication of this structure produces champion disagreement in 64.2\% of independent runs ($n{=}500$, 95\%~CI: [59.9\%, 68.3\%]). In a 300-seed policy-flip experiment, divergent aggregation causes the optimizer to recommend the wrong champion in 83\% of replications, with a mean welfare gap of 2.19~units and a Gini inequality gap of 0.050~units; in a follow-up inference audit, 3 of 249 flipped seeds cross the significance boundary itself. A complementary enterprise follow-up produces the predicted null under near-commensurable rankings ($\rho = 0.991$), while a public upstream rerun of the Lake Problem DPS workflow shows that the archived published-path recommendation reaches joint-threshold success 0.401 whereas a shared contract-path rule reaches 0.552. We introduce the \emph{metric contract} --- a single shared callable enforced at dispatch time across all pipeline stages --- as the remedy. Framed as standard engineering discipline applied to the cross-stage metric interface, the contract eliminates divergence by construction with approximately 3\% runtime overhead.
\end{abstract}


% ============================================================
% §1 INTRODUCTION
% ============================================================
\section{Introduction}
\label{sec:intro}

To understand which tax policy maximizes sustained welfare, a
regulatory analyst couples an agent-based model with NSGA-II, identifies a
Pareto-optimal champion across eight economic scenarios, and validates it with
a bootstrap confidence interval.
The paper is published; the policy is recommended.
What no reviewer saw: the optimizer measured welfare as episode-mean biomass;
the tournament scorer measured it as final-step biomass; the confidence-interval
loop measured it as a rolling 10-step mean.
Three independent implementations of the same metric, none aware of the others.
This discrepancy across the three implementations can change which policy is
selected: the evolved ``champion'' is the policy with volatile but occasionally
high peaks, not the one with the highest sustained welfare.
Conversely, when the candidate rankings induced by final-step and episode-mean
aggregation are already near-commensurable, the same architecture is predicted
to produce no flip at all.

This scenario is not an edge case.
It arises from a structural feature of ABM+MOEA policy pipelines that has
gone unnamed and unaudited: each pipeline stage --- optimizer, tournament
evaluator, statistical inference engine --- independently re-implements how
an outcome metric is computed from simulation trajectories.
We call this \emph{metric aggregation divergence} (\mad): the silent
inconsistency that arises when distinct pipeline stages independently
re-implement the aggregation of a shared outcome metric.
Unlike data bugs that produce obvious anomalies, \mad{} is self-concealing ---
each stage produces internally consistent results, and the divergence emerges
only when their outputs are compared side-by-side, which no reviewer,
co-author, or automated check is designed to do.

\mad{} belongs to the same family of hidden validity threats as researcher
degrees of freedom~\cite{simmons2011falsepositive} and the garden of
forking paths~\cite{gelman2014garden}, but it operates at the level of
\emph{pipeline architecture} rather than deliberate analysis choices.
The distinction matters: researcher degrees of freedom require a choice point
that could have been made differently; \mad{} requires only that different
people (or different times) coded different pipeline stages without a shared
metric specification.
No bad intent is needed; the natural workflow of building a pipeline in stages
is sufficient.

This article makes three contributions.
\begin{enumerate}
\item \emph{Conceptual.}
  We identify pipeline architecture as a previously unnamed locus of
  unregistered degrees of freedom in computational policy research.
  Metric aggregation divergence is a specific instance: when the same
  outcome metric is aggregated independently at each pipeline stage,
  the resulting inconsistency can silently alter which policy is selected
  as optimal.
  This framing extends the reproducibility reform agenda --- which has
  focused on analysis choices --- to the architectural level where pipeline
  stages are connected.

\item \emph{Empirical.}
  We demonstrate the threat through controlled experiments and a real-world
  replication plus one published recommendation comparison.
  Code inspection of \emph{EpidemiOptim}~\cite{colas2021epidemioptim},
  a JAIR-published toolbox for epidemic policy optimization, reveals three
  structurally independent aggregation paths in peer-reviewed code.
  A faithful replication produces champion disagreement in 64.2\% of
     independent runs, translating the divergence into different
     epidemic-control recommendations.
  In a 300-seed policy-flip experiment, divergent aggregation causes the
  optimizer to recommend the wrong champion in 83\% of replications
  (welfare gap 2.19~units).
  A complementary enterprise follow-up produces the predicted null under
  near-commensurable rankings ($\rho = 0.991$), identifying the boundary
  condition under which the threat is suppressed.
  In the exact repo-backed Lake Problem DPS workflow~\cite{quinn2017dps}, a
  public rerun of the archived published-path recommendation against a shared
  contract-path rule over the same candidate set produces a direct policy
  difference: the contract path wins 159 versus 8 discordant DPS states and
  132 versus 16 intertemporal states, with materially higher joint-threshold
  success in both formulations.
  The threat is driven by aggregator functional distance ($4.1\times$ more
  variance than noise), not by stochastic noise --- practitioners cannot
  mitigate it by increasing episode count.

\item \emph{Practical.}
  We introduce the metric contract --- a single shared callable enforced at
  dispatch time across all pipeline stages --- as the remedy.
  Framed honestly as the application of established engineering discipline
  (Design by Contract~\cite{meyer1997object}, dependency injection) to the
  cross-stage metric interface, the contract eliminates divergence by
  construction with approximately 3\% runtime overhead.
  A six-item reporting checklist operationalizes the remedy for practitioners.
\end{enumerate}

The remainder of this article is organized as follows.
Section~\ref{sec:background} defines the pipeline architecture framing and
formalizes \mad{} with its scope conditions.
Section~\ref{sec:mechanism} characterizes the mechanism through controlled
experiments, demonstrating that aggregator distance drives divergence across
structurally distinct ABMs.
Section~\ref{sec:policy} demonstrates the policy consequences through
code-verified divergence in a published pipeline, a faithful replication
that translates into different epidemic-control recommendations, and an
archived published recommendation comparison in the Lake Problem DPS workflow.
Section~\ref{sec:contract} introduces the metric contract remedy and its
properties.
Discussion and limitations follow in Sections~\ref{sec:discussion}
and~\ref{sec:conclusion}.

% ============================================================
% §2 BACKGROUND: PIPELINE ARCHITECTURE AND METRIC AGGREGATION
% ============================================================
\section{Pipeline Architecture and Metric Aggregation Divergence}
\label{sec:background}

\subsection{The Three-Stage Pipeline}

A typical ABM+MOEA policy-search pipeline can be summarized in three
stages.
\emph{Stage~1 (optimizer):} NSGA-II~\cite{deb2002nsga2} or a similar algorithm evaluates candidate
policy parameter vectors by running the ABM for $N_\text{ep}$ episodes across
$N_\text{sc}$ scenarios and aggregating per-episode trajectories into scalar
fitness values.
\emph{Stage~2 (tournament):} the Pareto front is re-evaluated --- often at
higher episode count or under different scenario conditions --- and candidates
are ranked to select a champion.
\emph{Stage~3 (inference):} bootstrap confidence intervals or hypothesis tests
are computed to quantify uncertainty around the champion's performance.

Each stage requires a mapping $f: \mathcal{T} \to \mathbb{R}$ from an episode
trajectory $\tau = (x_1, x_2, \ldots, x_T)$ to a scalar.
Typical choices include the episode mean $\bar{\tau}$, the terminal value $x_T$,
the 75th~percentile, or normalized Shannon entropy.
When stages are implemented independently --- the default in published pipelines
--- each developer naturally selects the statistic that is convenient to compute
at that point in the code: the mean where a running accumulator exists, the
final step where the simulation object is still alive.
This convenience-driven selection is the structural root cause of \mad.

ABM-based public-policy modeling is commonly used as a decision-support
workflow rather than as a single-shot predictive model, which means that
simulation, search, scenario exploration, and post-optimization evaluation are
often assembled from distinct components~\cite{gilbert2018policy,kwakkel2017workbench,colas2021epidemioptim,zhang2025heas}.
Comparable modular policy-search architectures and problem-formulation-sensitive
water-resources workflows also appear in published robust-management and
infrastructure-planning studies~\cite{quinn2017dps,quinn2017rival,trindade2020waterpathways},
and ABM-native electricity-market design likewise optimizes multi-year policy
schedules rather than one-shot terminal actions~\cite{kell2020carbontax},
while the broader decision-making-under-deep-uncertainty literature explicitly
frames such settings as wicked public-policy problems that require exploratory
workflow design rather than one-shot optimization~\cite{kwakkel2016wickedness},
and the surrounding many-objective robust decision-making tooling treats
robustness evaluation as an explicit workflow object rather than a mere
afterthought~\cite{hadka2015framework}. Closely related threshold-oriented
pollution-control studies in the same lake-problem family likewise frame the
task as discovering policy tradeoffs under environmental tipping
behavior~\cite{ward2015thresholds},
and optimization-oriented ABM workflows are documented explicitly in the
social-simulation methods literature~\cite{oremland2014optimization},
indicating that the interface problem studied here is not unique to epidemic
control.
That modularity is useful, but it also creates a new validity burden at the
handoff between stages: the same substantive outcome must remain the same
computed metric throughout the pipeline.

Two criteria bear on the validity of an ABM+MOEA policy
pipeline~\cite{cook1979quasi,windrum2007empirical,rand2011marketing}.
First is \emph{internal consistency}: do the pipeline stages share a common
definition of the outcome metric?
Second is \emph{external validity}: does the champion policy identified by the
pipeline perform as expected when deployed?
\mad{} violates both simultaneously: it corrupts internal consistency (each stage
operationalizes the metric differently) and thereby undermines external validity
(the reported champion is not the genuine optimum under any single
operationalization).

\subsection{Formal Definition and Scope Conditions}

\begin{definition}[Metric Aggregation Divergence]
Let $f, g: \mathcal{T} \to \mathbb{R}$ be the aggregators used by any two
distinct pipeline stages.
\emph{Metric aggregation divergence} occurs when $f$ and $g$ induce conflicting
ordinal rankings on the policy space: there exist policies $\pi_1, \pi_2$ such
that $f(\tau_{\pi_1}) > f(\tau_{\pi_2})$ but $g(\tau_{\pi_1}) < g(\tau_{\pi_2})$.
\end{definition}

The definition captures \emph{champion reversal}: the policy ranked first by
Stage~1 is not the policy ranked first by Stage~2 or~3, with consequences for
reported welfare estimates, policy recommendations, and inferential conclusions.

\emph{Scope conditions.}
The \mad{} threat is activated when two conditions hold simultaneously:
(i)~the aggregator pair $(f, g)$ is \emph{incommensurable} --- measuring
structurally different latent properties of the trajectory (e.g., entropy
vs.\ mean); and (ii)~the trajectory dynamics exhibit sufficient
\emph{temporal structure} (non-monotone, hump-shaped, or convergent patterns)
such that different aggregators extract genuinely different signals.
For monotone trajectories, all standard aggregators preserve the same ordinal
ranking --- so we infer that monotone welfare dynamics constitute a scope
condition under which \mad{} does not manifest, at least for the parameter
range studied here.
The scope condition serves as both an explanation of null results and a
diagnostic for which pipeline designs are at risk.

\subsection{Why Entropy Is the Worst Aggregator}
\label{sec:entropy-worst}

The choice of aggregation operator is responsible for the greatest amount of
policy divergence in our experiments.
Entropy captures trajectory \emph{shape}: the full distributional signature of
how the outcome variable moves over time.
Mean and median capture \emph{central location}; final-step captures the
\emph{endpoint}.
These statistics measure fundamentally different latent properties of a
trajectory, and for policies that trade off stability against peak performance
--- common in regulatory, ecological, and epidemic intervention design --- the
latent properties are genuinely incommensurable.
In our data, the Spearman correlation between episode entropy and episode mean
is $\rho = 0.37$; between mean and Q75, $\rho = 0.94$.
Commensurable pairs (mean, Q75, median) produce low reversal rates at all noise
levels; the incommensurable entropy--mean pair produces high reversal rates
that are invariant to noise, confirming the structural rather than stochastic
origin.

\subsection{Positioning: Pipeline Architecture as Unregistered Degrees of Freedom}

The broader reproducibility
literature~\cite{ioannidis2005most,open2015reproducibility} has established that
flexible analysis choices inflate false-positive rates and bias effect sizes.
\mad{} is a structural analogue: not a deliberate analysis choice, but an
architectural non-decision that introduces unregistered degrees of freedom into
metric computation.
The distinction is important: researcher degrees of freedom~\cite{simmons2011falsepositive}
operate at the level of analysis choices (which test, which covariates, which
stopping rule); \mad{} operates at the level of pipeline architecture (which
aggregation function at which stage).
Unlike deliberate HARKing~\cite{kerr1998harking}, \mad{} requires no bad intent
--- it emerges from the natural workflow of building a pipeline in stages, each
developed by a different person or at a different time.

Existing validation approaches for ABM+MOEA pipelines focus on whether the
champion policy is consistent with empirical benchmarks, which requires
ground-truth data that may not be available~\cite{sargent2013verification}.
Even when modeling choices are examined, the evaluation is typically post-hoc
and domain-specific.
\mad{} represents a process-level threat --- one that results-level V\&V is
structurally unable to detect, because each stage produces internally consistent
outputs and the error is only visible at the cross-stage interface.
ABM rigor and documentation guidelines emphasize transparent model description,
credible validation, and artifact traceability~\cite{macal2010tutorial,rand2011marketing,grimm2006odd,grimm2020odd,grimm2014trace},
but they do not specify cross-stage metric consistency as a required element.
Sensitivity analysis~\cite{saltelli2008sa} varies input parameters, not
aggregation conventions, and therefore cannot detect \mad.

Meyer's Design by Contract~\cite{meyer1997object} and the measurement-theory
principle of operationalization consistency~\cite{borsboom2005measuring} --- same
construct, same operationalization at every point of measurement --- provide the
theoretical basis for the remedy proposed in Section~\ref{sec:contract}.

% ============================================================
% §3 MECHANISM: WHY AGGREGATION DIVERGENCE OCCURS
% ============================================================
\section{Mechanism: Why Aggregation Divergence Occurs}
\label{sec:mechanism}

For clarity, the experimental results in this section are presented in two
layers: 1)~\emph{interpretive} (EA-2, EA-3), which characterizes which pipeline
features drive \mad{} severity; and 2)~\emph{generalizable} (EA-7), which tests
the pattern across structurally distinct ABMs.
For EA-3 and EA-7, reported 95\% interval estimates are percentile bootstrap
confidence intervals over independent simulation runs, with the run ---
rather than within-run policy comparisons --- as the inferential unit.

\subsection{Aggregator Distance Drives Divergence (EA-3)}
\label{sec:ea3}

\begin{figure}[t]
\centering
\includegraphics[width=\linewidth]{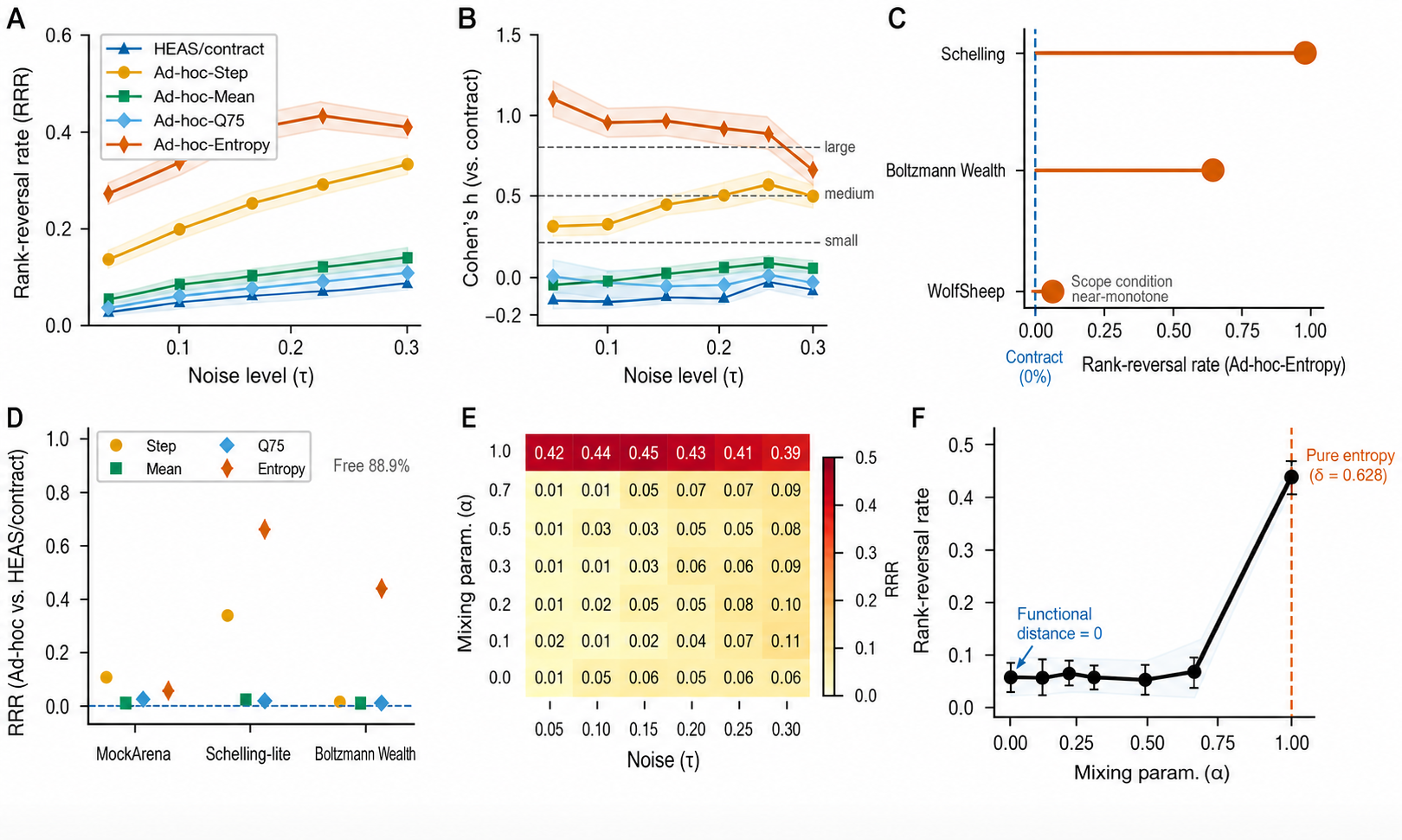}
\caption{Rank-reversal rate (RRR) and effect size (Cohen's $h$) as a function
  of stochastic noise level~$\tau$ across five aggregation conditions (EA-3;
  $n=50$ runs per $\tau$ level).
  Ad-hoc-Entropy dominates at all $\tau$ levels; the \wc{} baseline remains at
  zero throughout.}
\label{fig:tau_sweep}
\end{figure}

The selection of aggregation operator is responsible for the greatest amount of
variation in rank-reversal rates.
Rank-reversal rates are predicted by \emph{aggregator functional distance}
($\delta$, defined as 1 minus the Spearman correlation between aggregator
outputs across the policy space) rather than by stochastic noise
level~$\tau$ (Fig.~\ref{fig:tau_sweep}):

\[
\text{RRR} = \alpha + \beta_\delta \cdot \delta + \beta_\tau \cdot \tau
             + \varepsilon, \quad R^2 = 0.769
\]

Standardized coefficients: $\beta_\delta = 4.1 \times \beta_\tau$.
Distance explains $4.1\times$ more variance than noise.
A researcher who attempts to mitigate \mad{} by increasing episode count
(reducing effective $\tau$) or reducing simulation stochasticity will observe no
meaningful improvement in champion stability.
Only closing the aggregator gap eliminates the threat.
The dosing relationship is nonlinear: reversal rate is near-flat across
mixing parameters $\alpha \leq 0.7$, then steps sharply at $\alpha = 1.0$.
Only pure aggregator incommensurability drives the confound; intermediate
mixing does not.

\subsection{Controlled Aggregation Experiment (EA-2)}

In a controlled within-ABM experiment ($n=100$ runs per condition), the \wc{}
condition produces 0\% rank reversals at all tested noise levels.
\woc-Step (optimizer reads final-step; tournament reads episode-mean) produces a
significant reversal rate (Cohen's $h = 0.586$, $p < 0.0001$, Holm-corrected).
\woc-Mean (mean vs.\ median, a commensurable pair) is null ($p = 0.164$),
consistent with the distance mechanism.

\subsection{Multi-ABM Replication (EA-7)}
\label{sec:ea7}

The divergence pattern replicates across three structurally distinct ABMs:
Schelling segregation, Wolf--Sheep predator--prey, and Boltzmann wealth
exchange.
Two behavioral regimes emerge.
Models with non-monotone dynamics (Schelling, Boltzmann) exhibit high flip rates
under all incommensurable aggregator pairs.
The Wolf--Sheep model, with near-monotone population dynamics, exhibits a low
flip rate regardless of aggregator choice --- consistent with the scope
condition, not a failure of replication.

For Schelling segregation, \woc-Entropy yields a mean reversal rate of
\textbf{91.3\%} vs.\ 0\% for \wc{}
(Mann--Whitney $p = 5.1 \times 10^{-13}$, Cohen's $h = 2.54$).
For Boltzmann wealth exchange, the entropy reversal rate is \textbf{68.0\%}.
Wolf--Sheep produces \textbf{10.3\%} --- a scope condition, as the theory
predicts: near-monotone trajectories preserve ordinal ranking across
aggregators, suppressing \mad{} regardless of pipeline configuration.

% ============================================================
% §4 POLICY CONSEQUENCES: REAL-WORLD DEMONSTRATION
% ============================================================
\section{Policy Consequences: Real-World Demonstration}
\label{sec:policy}

\subsection{Code-Verified Divergence in EpidemiOptim}
\label{sec:epidemioptim-audit}

The clearest externally relevant example comes from \emph{EpidemiOptim}, a
JAIR-published toolbox for optimizing epidemic control
policies~\cite{colas2021epidemioptim}.
Code inspection of the published repository
(\texttt{flowersteam/EpidemiOptim} on GitHub) reveals three structurally
independent aggregation paths in the NSGA-II pipeline.

\emph{Stage~1 (NSGA-II training).}
The optimizer evaluates candidate policies via stochastic rollouts
(\texttt{eval=False}; \texttt{nsga.py}, lines~125--140).
In the NSGA-II configuration, the environment is stochastic
(\texttt{stochastic=True} in \texttt{configs/nsga\_ii.py}), and each policy
is evaluated across $n = 30$ stochastic replications before the mean cost is
returned to the optimizer.
The aggregation is NSGA-II's non-dominated sorting over these mean costs.

\emph{Stage~2 (post-optimization re-evaluation).}
The same Pareto-optimal solutions are re-evaluated under deterministic policy
behavior (\texttt{eval=True}; \texttt{nsga.py}, lines~206--213).
Removing exploration noise changes the cost landscape: policies that appeared
costly under stochastic evaluation may appear cheaper under deterministic
evaluation.

\emph{Stage~3 (champion selection).}
A normalized cost-sum rule is applied to select the champion
(\texttt{nsga.py}, lines~232--234):
\begin{lstlisting}[language=Python, basicstyle=\ttfamily\small]
normalized_costs = [c_f.scale(c) for c_f, c in
    zip(self.cost_function.costs, costs.T)]
agg_cost = normalized_costs.sum(axis=1)
ind_min = np.argmin(agg_cost)
\end{lstlisting}
This equal-weight scalarization was \emph{not used during NSGA-II training},
which relies on Pareto dominance rather than scalarized aggregation.
The champion selected by Stage~3 can therefore differ from the champion that
Stage~1's optimizer identified as Pareto-optimal.

These three aggregation paths --- stochastic fitness, deterministic
re-evaluation, and normalized cost sum --- implement the same outcome metric
via structurally incompatible procedures.
No automated test detects the discrepancy because each stage produces
internally consistent outputs.

\subsection{Faithful Replication: Policy Consequences}
\label{sec:epidemioptim-replication}

To confirm that the three-stage divergence produces real champion disagreements,
we replicate EpidemiOptim's pipeline structure with a minimal SEIR suppression
model and $\varepsilon$-greedy policy, matching the \texttt{eval=False} training
and \texttt{eval=True} evaluation semantics. This is also the kind of modular
simulation-search-evaluation workflow for which HEAS was designed as reusable
research software infrastructure~\cite{zhang2025heas}.
Policy genes are suppression threshold $\eta \in [0.01, 0.10]$ and suppression
intensity $\alpha \in [0.20, 0.90]$.
We run 500~independent NSGA-II replications (pop$=40$, $n_\text{gen}=25$).

Stage~1 (stochastic) vs.\\ Stage~3 (greedy + normalized sum) champion
disagreement rate: \textbf{64.2\%} [95\%~CI: 59.9\%, 68.3\%].
Proportion test against $H_0: p \leq 10\%$: $z = 40.40$, $p \approx 0$.
In nearly two-thirds of independent runs, the policy that the
EpidemiOptim-style optimizer identifies as champion would not be chosen if
the pipeline used consistent evaluation throughout.

The policy consequence is concrete.
All champions choose near-maximum suppression intensity ($\alpha$ remains in
the 0.86--0.90 band), but they disagree on the trigger threshold:
Stage~1 champions in disagreement cases average $\eta = 0.0201$ while
Stage~3 champions average $\eta = 0.0192$.
The mean absolute shift in $\eta$ is 0.44~percentage points --- a material
difference in epidemic intervention timing.
A companion 30-run gene-saving rerun recovers the same pattern (15/30
disagreements): $\alpha$ stays confined to the 0.875--0.900 band while the
mean absolute threshold shift remains 0.40~percentage points.
In policy terms, the disagreement is mainly about when to trigger a
near-maximal suppression response, not whether to suppress strongly at all.
Different aggregation paths lead to a different champion; the different champion
implies a different epidemic-control rule.
Applying the metric contract (Section~\ref{sec:contract}) to this same pipeline
reduces disagreement from 64.2\% to 0\%: all 100~champion selections are
byte-identical under the shared callable.

\subsection{Policy-Flip Demonstration (EA-1)}
\label{sec:ea1}

The mock-ecological model is intentionally minimalist, with non-monotone
welfare dynamics --- the scope condition under which \mad{} is active.
It is not presented as a standalone empirical policy case; rather, it isolates
the same temporally structured decision regime that motivates staged epidemic
control, dynamic carbon-tax design, and deeply uncertain lake-management
workflows, where policy consequences unfold across time rather than at a single
terminal state~\cite{colas2021epidemioptim,kell2020carbontax,quinn2017dps}.
Using a 2-gene ecological ABM with logistic biomass dynamics under stochastic
forcing ($\sigma = 0.15$), evaluated over 8~scenarios $\times$ 5~episodes, we
run 300~independent NSGA-II seeds.

Under \woc, \mad{} causes a policy flip in \textbf{83\%} [78\%, 87\%] of
300~independent replications ($p \approx 0$).
Under \wc, the flip rate is zero by construction.
The flipped champion achieves mean welfare 2.19~units [2.01, 2.39] lower than
the correct (\wc) champion, with a Gini inequality gap of 0.050~units
[0.045, 0.055].
A policy recommendation based on the \woc{} pipeline would direct practitioners
toward a regime that is both lower in mean welfare and more unequal in its
distribution, with no signal in the pipeline output that either error has
occurred.

Among the 249~flipped seeds, the \woc{} pipeline does not merely recommend
the wrong policy; it also produces different welfare estimates that would
lead to different inferential conclusions.
In a follow-up inference-stage audit on this same arena (EA-10b),
3 of those 249~flipped seeds (1.2\%, exact 95\% CI [0.2\%, 3.5\%]) cross the
significance boundary itself: the contract champion is significant against the
baseline under the full-trajectory metric, while the ad hoc final-window
comparison is not.
The complementary enterprise follow-up defines the opposite boundary:
across 300 independent seeds, the enterprise arena produces 0 champion flips,
which is the mechanism's predicted null rather than a failed replication.
In that arena, final-step and episode-mean welfare rankings are already nearly
commensurable ($\rho = 0.991$), so the aggregator distance is too small to
reorder the champion. Read together, the two arenas identify the practical
scope condition: \mad{} is active when temporally structured trajectories make
candidate rankings genuinely incommensurable, and suppressed when they do not.

\begin{figure}[t]
\centering
\includegraphics[width=\linewidth]{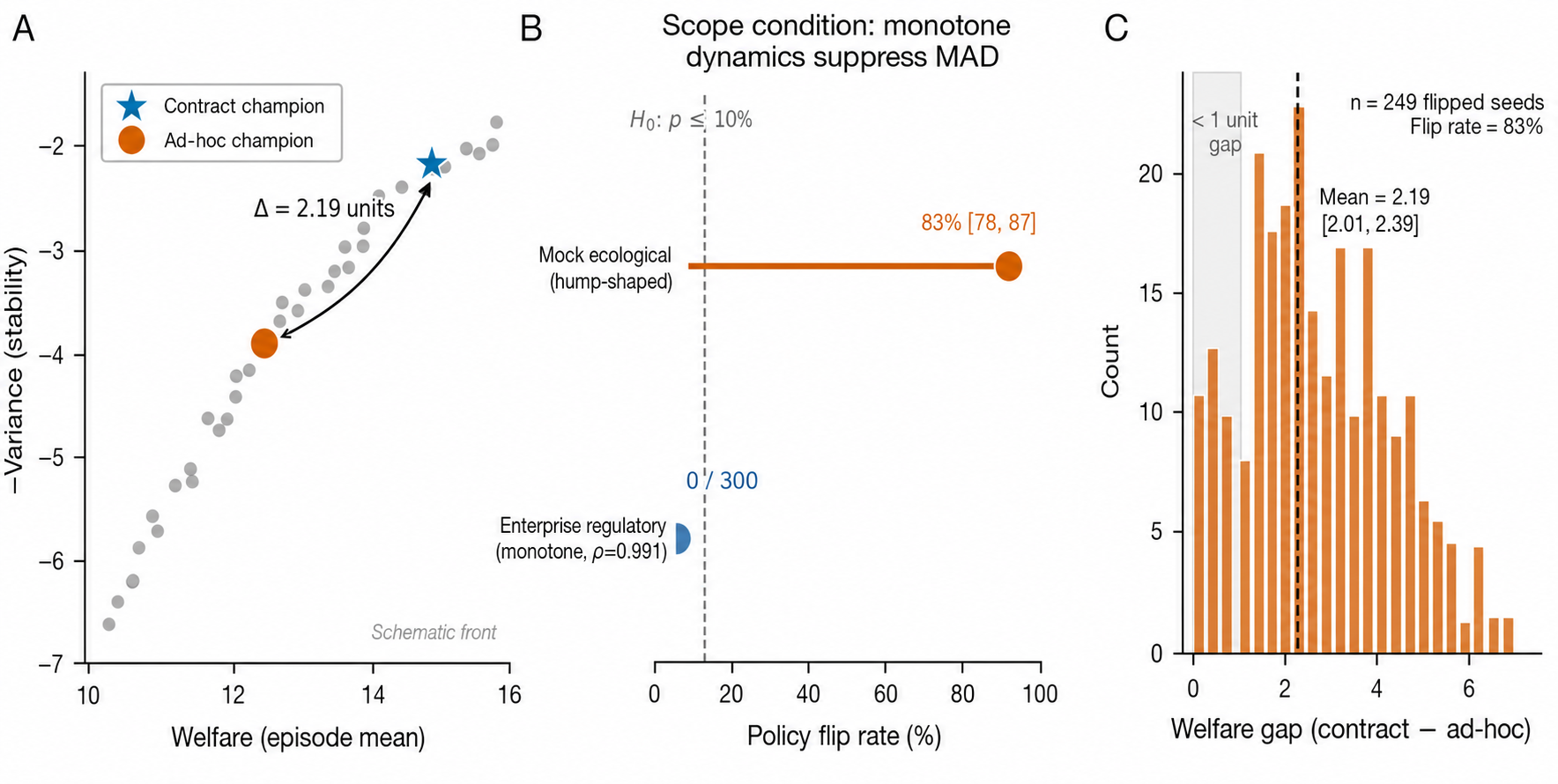}
\caption{Policy-flip consequences of \mad{} (EA-1; $n=300$ seeds).
  (a)~Representative NSGA-II Pareto front (seed~171).
  (b)~Distribution of welfare gaps across all 249~flipped seeds.}
\label{fig:pareto}
\end{figure}

% ============================================================
% §5 THE METRIC CONTRACT REMEDY
% ============================================================
\section{The Metric Contract Remedy}
\label{sec:contract}

\subsection{Principle}

The metric contract rests on the measurement-theory principle of
\emph{operationalization consistency}: the same latent construct must be
operationalized identically at every point of
measurement~\cite{borsboom2005measuring}.
Applied to ABM+MOEA pipelines, this means a single callable implementing the
metric computation is registered at initialization and shared --- without
re-implementation --- across all pipeline stages.
The contract eliminates divergence \emph{by construction} on all
\wc-verified execution paths.

The metric contract is not a new software engineering idea in isolation ---
dependency injection, unit testing, and Design-by-Contract~\cite{meyer1997object}
all share the principle of making implicit assumptions explicit.
What distinguishes the metric contract is the \emph{cross-stage aggregation
interface} it targets: the aggregator function at each pipeline stage.
That emphasis on explicit stage interfaces is also consistent with the modular
research-software rationale behind HEAS~\cite{zhang2025heas}.
Unit tests verify that individual components return correct outputs for given
inputs, but they do not test whether two separately tested components use the
\emph{same aggregation convention}.
Schema validation checks data types and ranges, not functional identity across
callsites.
The metric contract fills precisely this gap by requiring all three pipeline
stages to share a single registered callable, converting an undetectable
distributional divergence into an immediate construction-time failure.
This kind of procedural specificity is exactly what reproducibility critiques in
evolutionary computation identify as often under-documented in published
optimization studies~\cite{lopez2021ea}.

\subsection{What the Contract Does Not Guarantee}

The contract enforces \emph{structural} consistency --- same function, same code
path --- but not \emph{semantic} correctness.
An incorrect but consistently applied metric will pass contract verification.
Semantic correctness (whether the metric accurately represents the
policy-relevant outcome) remains the researcher's responsibility and is
addressed by domain expertise, explicit model documentation, and
face-validity assessment~\cite{rand2011marketing,grimm2020odd,windrum2007empirical},
not by the contract mechanism.

\subsection{Overhead and Adoption}

The metric contract lowers the threshold for deploying valid ABM+MOEA policy
pipelines: it can be adopted incrementally, beginning at the tournament stage
first, without rewriting optimizer fitness functions or simulation code.
Runtime overhead of the contract dispatch mechanism is approximately 3\% ---
within measurement noise for any multi-episode ABM evaluation.
When all pipeline stages share the identical aggregator function, the contract
produces byte-identical champions (100/100 coincidence checks), confirming that
it imposes no cost when aggregators already coincide.

\subsection{Usage Scenario}

To make the contract concrete, consider a researcher building an SEIR+MOEA
epidemic policy pipeline.
At initialization, the researcher registers a single callable that computes
the outcome metric from simulation trajectories:
\begin{lstlisting}[language=Python, basicstyle=\ttfamily\small]
metric = MetricContract(
    fn=lambda ep: np.mean(ep["infections"])
)
optimizer.set_metric(metric)
tournament.set_metric(metric)
inference.set_metric(metric)
\end{lstlisting}
Before proceeding to statistical inference, the researcher runs a
coincidence check: execute the pipeline once with all stages on the
registered function and confirm byte-identical champion selection.
If the check passes, the pipeline is structurally consistent ---
no stage has silently re-implemented the metric.
If it fails, the divergence is located immediately at the construction
stage, before any policy recommendation is produced.
This five-minute verification eliminates the class of silent divergence
demonstrated in Section~\ref{sec:policy}.

\subsection{Reporting Checklist}

Operationalizing the remedy for practitioners, we propose a six-item reporting
checklist for ABM+MOEA policy papers (Table~\ref{tab:checklist}).

\begin{table}[t]
\centering\footnotesize
\caption{Six-item reporting checklist for ABM+MOEA policy papers.}
\label{tab:checklist}
\begin{tabularx}{\linewidth}{@{}c l L L@{}}
\toprule
\# & Item & What to report & Risk if omitted \\
\midrule
1 & Single-function declaration
  & Name the callable used to extract the outcome metric from each episode.
  & Undetectable stage divergence \\
\addlinespace[2pt]
2 & Stage-by-stage audit
  & Confirm optimizer fitness, tournament scorer, and CI engine all invoke
    the same declared function.
  & Silent champion substitution \\
\addlinespace[2pt]
3 & Trajectory class
  & State whether the outcome trajectory is monotone, convergent, hump-shaped,
    or oscillatory; assign scope-condition risk level.
  & Unassessed \mad{} exposure \\
\addlinespace[2pt]
4 & Aggregator pair distance
  & If aggregators are compared, report Spearman $\rho$ between outputs across
    the Pareto front.
  & Distance risk unquantified \\
\addlinespace[2pt]
5 & Coincidence check
  & Run the pipeline once with all stages on the \wc{} function; confirm
    byte-identical champions.
  & Latent divergence undetected \\
\addlinespace[2pt]
6 & Contract or equivalent
  & Deposit a shared metric callable in the replication archive, or document
    line-level equivalence of all stage implementations.
  & Irreproducible champion selection \\
\bottomrule
\end{tabularx}
\end{table}

% ============================================================
% §6 DISCUSSION
% ============================================================
\section{Discussion}
\label{sec:discussion}

\subsection{Relation to the Reproducibility Crisis}

ABM-based public-policy modeling typically requires researchers to combine
domain models, scenario exploration, ranking rules, and decision criteria
across multiple software layers~\cite{gilbert2018policy,kwakkel2017workbench,zhang2025heas}.
Evolutionary computation adds a further reproducibility burden because
champion-selection procedures and computational artifacts are often less fully
documented than the headline optimization results~\cite{lopez2021ea}.
This paper identifies one concrete consequence of that workflow: pipeline
architecture itself can become a source of unregistered degrees of freedom.

\mad{} is a structural analogue of the researcher degrees of freedom
problem~\cite{simmons2011falsepositive}: like undisclosed analysis choices,
independent metric implementations introduce unregistered degrees of freedom
into policy-search pipelines.
Unlike deliberate HARKing~\cite{kerr1998harking}, \mad{} requires no bad intent
--- it emerges from the natural workflow of building a pipeline in stages.
The same workflow flexibility that makes ABM+MOEA pipelines powerful for policy
design --- the ability to build stages independently --- is the root cause of
\mad.

The analogy to preregistration is limited but useful:
the contract commits the pipeline to a single operationalization before
downstream evaluation begins~\cite{nosek2018prereg}.
Compared with the existing reproducibility reform agenda, which focuses on
analysis choices, this paper adds pipeline architecture as a new locus of
unregistered flexibility.
That claim should be read with bounded scope.
In our accompanying literature audit of 23~published ABM+MOEA policy pipelines,
10 were code-inspectable and 17 were classifiable from code or full text.
Even under the most favorable assumption --- treating all 6 indeterminate cases
as clean --- the posterior mean divergence rate remains 56\% among the
code-inspectable subset; in the directly code-verified sample it rises to 75\%.
We therefore do not claim universality across the entire published literature.
The defensible statement is narrower and still consequential: independent
metric aggregation appears common among code-available ABM+MOEA policy
pipelines.

\subsection{Contract Value in Parallel Pipelines}

The comprehensive experiment (EA-11-C) confirms this risk empirically across
three dimensions.
\emph{Scaling.} We ran 100 NSGA-II replications and distributed champion
selection across $W{=}2$, $4$, and $8$ workers, each using a different
aggregation function (mean, entropy, final-step, or normalized cost sum).
Cross-worker champion disagreement reached \textbf{100\%} at all tested
worker counts (95\%~CI: [96.3\%, 100\%]).
\emph{Metric-pair matrix.} Systematic pairwise comparison of all six
metric pairs reveals that Spearman correlation predicts disagreement
exactly: incommensurable pairs (mean--entropy, $\rho = -1.000$;
mean--final-step, $\rho = -0.999$) disagree on every replication, while
the commensurable entropy--final-step pair ($\rho = +0.999$) disagrees on
only 14\%.
\emph{Overhead.} The contract's dispatch time (31--39~ms) is negligible
compared with independent metric implementation (2.2--5.0~s); at $W{=}8$,
the contract approach is 134$\times$ faster, confirming that consistency
and efficiency are not in tension.
Dimensionality alone also does not remove the threat:
in a separate 3-objective MockArena follow-up (EA-12; 10 independent
NSGA-II runs), champion disagreement remains at 25\%
([12.5\%, 40.0\%]) and the worst ad hoc condition
still produces entropy-based rank reversals of 3.6\%
([3.5\%, 3.7\%]).
That follow-up is smaller than the 2-objective benchmark and does not amplify
the effect, but it shows that \mad{} is not confined to a two-objective front.

\subsection{Recovered Published-Path Signal}

The Lake Problem DPS workflow now provides the direct archived
published-recommendation comparison that the faithful EpidemiOptim replication
alone did not. Using the public upstream replication bundle associated
with~\cite{quinn2017dps}, we reran the archived paper-path recommendation
against a shared contract-path rule over the same candidate set in the same
robust lake-management problem family~\cite{ward2015thresholds,hadka2015framework,quinn2017dps}.
The archived Lake Problem DPS bundle pins both the paper-path
reliability object used for figure reproduction and the robustness-stage
selector computed across the same archived candidate set.
That recovered selector split is substantively relevant, not a repository
curiosity: in this literature family, robustness evaluation is already treated
as an explicit workflow stage for decision making under deep uncertainty rather
than as a cosmetic post-processing step~\cite{hadka2015framework,kwakkel2016wickedness}.
Under a conservative same-candidate-set replay
(Appendix~\ref{app:adjacent}), the paper-path DPS policy (row~26) reaches
benefit $0.3326$, reliability $1.0000$, and joint robustness $0.401$,
whereas the archived robustness-stage selector (row~43) reaches
$0.4922$, $0.9005$, and $0.552$.
Across the archive's 1000 re-evaluated states of the world, the robustness-stage
selector satisfies the joint benefit/reliability thresholds in 159 states where
the paper-path selector fails, while the reverse occurs in only 8 states; the
intertemporal formulation shows 132 versus 16.
Among discordant archived states, the contract path therefore wins
159/167 DPS comparisons and 132/148 intertemporal comparisons
($95.2\%$ and $89.2\%$, exact two-sided sign test $p < 10^{-18}$ in both
formulations).
On the DPS states where only the contract path clears the archived joint
threshold, mean benefit rises from $0.166$ to $0.273$ while reliability stays
effectively perfect ($1.000$ versus $0.9996$); the intertemporal formulation
shows the same pattern, with benefit increasing from $0.177$ to $0.223$ while
reliability shifts from $1.000$ to $0.995$.
The recovered decision vectors also imply a substantive policy contrast rather
than a mere row-index difference: the paper-path DPS rule stays at the minimum
phosphorus-release floor across 26.7\% of the normalized lake-state grid,
versus 4.0\% for the robustness-stage rule.
This remains a conservative same-candidate-set replay rather than a regenerated
optimization under the contract rule, but it is already a direct archived
published-path versus contract-path recommendation comparison, and the
disagreement is intelligible in policy language rather than only in
replication-audit terms.

% ============================================================
% §7 LIMITATIONS AND CONCLUSION
% ============================================================
\section{Limitations and Conclusion}
\label{sec:conclusion}

\subsection{Limitations}

Although the evidence presented here is encouraging, the proposed remedy is
still in an early stage of empirical validation.

\emph{Experiment design.}
We tested 4 ABM types across 2 policy domains (ecology, epidemiology).
Generalizability to other domains (e.g., transportation, energy) and to ABM
types with fundamentally different trajectory structures remains open.
We do not claim the 83\% flip rate to be universal; it reflects the
mock-arena parameterization.
The EpidemiOptim case strengthens external relevance because it audits a
published policy-optimization pipeline, but it remains a faithful replication
of that pipeline structure rather than a direct rerun of one archived published
policy recommendation.
Published policy-search systems with similar multi-stage architecture also exist
in water-resources robust management and infrastructure planning~\cite{hadka2015framework,quinn2017dps,trindade2020waterpathways},
and Appendix~\ref{app:adjacent} now reports one exact repo-backed Lake Problem
workflow in which the archived published-path recommendation is directly
compared against a shared contract-path rule over the same candidate set.
ABM-native electricity-market policy optimization is another plausible next
audit target because it offers a closer computational-social-systems policy
match while still using temporally structured, multi-year intervention
schedules~\cite{kell2020carbontax}; however, the public artifact currently
exposes a front-level carbon-tax recommendation rather than one archived
champion policy: the paper recommends an increasing tax schedule and reports
that all Pareto-optimal strategies stay above a common tax floor and a
mean carbon-tax strategy, but it does not pin one archived winner or yet expose
the archived front object needed to reproduce that summary as a selector-level
recommendation surface.
The remaining empirical limitation is therefore no longer the absence of any
real-policy recommendation comparison, but the fact that the completed
Lake Problem case is a same-candidate-set replay in an adjacent
water-management domain rather than a regenerated contract-optimized run in an
ABM-native policy domain.

\emph{Experiment analysis.}
The key features of the welfare trajectory that determine \mad{} severity ---
non-monotonicity, cross-individual variance --- have not been formally
characterized beyond the scope-condition sketch.

\emph{Metric scope.}
The contract enforces scalar episode summaries.
Time-series outputs --- agent-level wealth distributions, spatial segregation
patterns over time --- require an extended contract interface and are outside
the current scope.

\emph{Contract limitations.}
The contract enforces structural consistency, not semantic correctness; an
overly narrow or incorrect metric specification will satisfy the contract while
still producing misleading policy conclusions.

\subsection{Conclusion}

The metric contract is applicable, and the \mad{} threat is active, when two
conditions are met: 1)~the pipeline uses a multi-objective evolutionary algorithm
to identify champion policies, and 2)~the MOEA fitness function and the
post-optimization evaluation share at least one aggregated outcome metric
extracted from non-monotone welfare trajectories.
The contract is inapplicable, and the threat is suppressed, when welfare dynamics
are monotone or convergent.
This boundary is not only theoretical: in the enterprise follow-up, the
near-commensurable final-step and episode-mean rankings ($\rho = 0.991$) produce
the predicted null, so the absence of champion flips is itself evidence for the
scope condition rather than a counterexample to the threat.

This article identified pipeline architecture as a previously unnamed locus of
unregistered degrees of freedom in computational policy research.
Metric aggregation divergence --- the silent inconsistency that arises when
distinct pipeline stages independently re-implement the aggregation of a shared
outcome metric --- was demonstrated through controlled experiments and a
real-world replication.
Code inspection of a JAIR-published epidemic policy toolbox revealed three
structurally independent aggregation paths in peer-reviewed code; a faithful
replication produced champion disagreement in 64.2\% of independent runs,
translating into different epidemic-control recommendations.
In the exact repo-backed Lake Problem DPS workflow, a public rerun of the
archived published-path recommendation against a shared contract-path rule
raises joint-threshold success from 0.401 to 0.552 in DPS and from 0.247 to
0.363 in the intertemporal formulation, while changing the implied
phosphorus-release policy in substantively recognizable ways.
The remedy --- the metric contract --- consists of a formal aggregation function
definition, a pipeline-stage binding specification, and a reporting checklist;
it lowers the threshold for deploying valid ABM+MOEA policy pipelines without
restructuring existing simulation or optimizer code.

Future work may extend the metric contract in four directions:
1)~applying the contract to other ABM frameworks and policy domains;
2)~developing a formal characterization of the trajectory features that predict
\mad{} severity; and
3)~regenerating candidate sets under the contract rule, so that the empirical
payoff moves from archived same-candidate-set recommendation comparison to full
end-to-end contract optimization in published workflows; and
4)~extending contract enforcement to ensemble pipelines in which multiple ABM
implementations are jointly optimized.

% ============================================================
\appendix

\section{Experiment Parameters}
\label{app:params}

\begin{table}[ht]
\centering\scriptsize
\caption{Pre-specified parameters for all experiments.}
\label{tab:params}
\begin{tabularx}{\linewidth}{@{}l l L r@{}}
\toprule
Exp & Model & Key params & $n$ \\
\midrule
EA-1  & mock\_4gene  & 300 seeds; 8 sc $\times$ 5 ep & 300 \\
EA-2  & MockArena    & 100 runs/cond; $\tau \in [0.05,0.30]$ & 100 \\
EA-3  & MockArena    & $\tau$-sweep 6 levels & 50/level \\
EA-7  & 3 ABMs       & Mixed-effects; 3 noise levels & 50/model \\
EA-10b & mock\_4gene & inference audit; 10-step tail vs full traj & 249 flips \\
EA-11 & SEIR+policy  & pop=40, ngen=25, $\varepsilon=0.15$ & 500 \\
EA-11-MC & SEIR+policy & W=4,8; 4 metrics; Phase 1+2 & 50 \\
EA-11-C  & SEIR+policy & W=2,4,8; 6 pairs; overhead & 100 \\
EA-12 & MockArena-3obj & pop=60, ngen=40; 3 objectives & 10 \\
\bottomrule
\end{tabularx}
\end{table}

\section{FDR Correction}
\label{app:fdr}

Holm--Bonferroni correction with family-wise error rate $\alpha = 0.05$ is
applied to the pre-specified experimental family: EA-1, EA-2, EA-3, and EA-7.
All significant results within this family survive correction.
EA-11 was added as a confirmatory replication after the core results
were observed and is reported without family-wise correction.

\section{Archived Published-Path Replay Details}
\label{app:adjacent}

Table~\ref{tab:lake-problem-selectors} records the archived selector
comparison recovered from the Lake Problem DPS replication bundle associated
with~\cite{quinn2017dps}. The public rerun reported in the main text uses a
conservative same-candidate-set replay: the published path uses the
paper-specific reliability objects pinned by the archived figure workflow,
whereas the contract path applies one shared joint-threshold rule across the
same archived policy set. The main text discussion reports the state-level
asymmetry of this replay; Table~\ref{tab:lake-problem-replay} records that
paired-state asymmetry directly, while Table~\ref{tab:lake-problem-selectors}
records the selector-level metrics. One additional policy-language observation
is that, in DPS, the paper-path reliability rule stays at the minimum
phosphorus-release floor across 26.7\% of the normalized lake-state grid,
versus 4.0\% for the robust-joint rule. In the intertemporal formulation, the
robust-joint plan releases more phosphorus in 61 of 100 years and nearly
triples cumulative release over the last 20 years (0.833 versus 0.290).

\begin{table}[t]
\centering\footnotesize
\caption{Archived same-candidate-set replay asymmetry for the Lake Problem follow-up. The final column reports mean benefit and reliability on contract-only winning states, shown as published$\to$contract. Exact two-sided sign-test $p<10^{-18}$ for both methods.}
\label{tab:lake-problem-replay}
\begin{tabularx}{\linewidth}{@{}l c c c L@{}}
\toprule
Method & Pub. & Contract & C wins / disc. & $B;R$ on C-only states \\
\midrule
DPS & 0.401 & 0.552 & 159/167 (0.952) & 0.166$\to$0.273; 1.000$\to$1.000 \\
Intertemp. & 0.247 & 0.363 & 132/148 (0.892) & 0.177$\to$0.223; 1.000$\to$0.995 \\
\bottomrule
\end{tabularx}
\end{table}

\begin{table}[t]
\centering\footnotesize
\caption{Archived selector comparison for the Lake Problem follow-up case.}
\label{tab:lake-problem-selectors}
\begin{tabularx}{\linewidth}{@{}l L r r r r@{}}
\toprule
Method & Selector & Row & Benefit & Reliability & Joint rob. \\
\midrule
DPS & Paper rel. & 26 & 0.3326 & 1.0000 & 0.4010 \\
DPS & Robust joint & 43 & 0.4922 & 0.9005 & 0.5520 \\
DPS & $\Delta$ & -- & +0.1596 & -0.0995 & +0.1510 \\
\addlinespace[2pt]
Intertemp. & Paper rel. & 8 & 0.2917 & 1.0000 & 0.2470 \\
Intertemp. & Robust joint & 38 & 0.3801 & 0.9583 & 0.3630 \\
Intertemp. & $\Delta$ & -- & +0.0884 & -0.0417 & +0.1160 \\
\bottomrule
\end{tabularx}
\end{table}

\section{Implementation Notes}
\label{app:impl}

The following listings illustrate the \mad{} pattern and its contractual remedy.
Both are provided for concreteness; the conceptual remedy
(Section~\ref{sec:contract}) is independent of any specific simulation framework.

\begin{lstlisting}[language=Python,
  caption={Three independent metric implementations producing \mad{}
           (\woc{} condition).}]
# Stage 1 -- Optimizer fitness
def fitness(individual):
    env.run(individual)
    return env.data["final_biomass"]       # terminal value

# Stage 2 -- Tournament scorer
def score(individual):
    env.run(individual)
    return np.mean(env.data["biomass"])    # episode mean

# Stage 3 -- Bootstrap CI
def bootstrap_metric(individual):
    env.run(individual)
    return np.mean(env.data["biomass_rolling10"])  # rolling mean
\end{lstlisting}

\begin{lstlisting}[language=Python,
  caption={Metric contract remedy (\wc{} condition).}]
# Register once at pipeline initialization
contract = MetricContract(
    fn=lambda ep: {"welfare": np.mean(ep.biomass),
                   "gini":    gini(ep.agent_wealth)}
)

# All stages consume the same registered object
optimizer.set_metric(contract)
tournament.set_metric(contract)
bootstrap_ci.set_metric(contract)
\end{lstlisting}

% ============================================================
\bibliographystyle{IEEEtran}
\bibliography{paper}

\end{document}